\documentclass[12pt]{article}
\usepackage[dvips]{graphicx}
\usepackage{amsmath,amsfonts,amssymb}
\usepackage[mathscr]{eucal}
\usepackage{verbatim}
\usepackage{epsfig}

\def\beq{\begin{equation}}
\def\eeq{\end{equation}}
\def\beqn{\begin{eqnarray}}
\def\eeqn{\end{eqnarray}}

\let\pes\, \let\esm\: \let\esn\ \let\ges\;

\def\a{\alpha} \def\b{\beta}

\def\o{\omega}

\def\L{\Lambda}

\def\bs{\boldsymbol{\sigma}}

\let\rel=\relatif 

\def\Reel{\mathop{\rm I\! R}}

\title{\Large \bf
Euler--Poincar\'e Characteristic and\\ Phase Transition in the Potts Model}

\author{\normalsize \sc Philippe Blanchard$^{1}$, Santo Fortunato$^{2}$
and Daniel Gandolfo$^{1,3,4}$ }

\date{}

\begin{document}

%\draft

\maketitle
\bibliographystyle{plain}

\footnotetext[1]{ Fakult\"{a}t f\"{u}r Physik, Theoretische Physik and BiBos,
Universit\"{a}t Bielefeld, Universit\"{a}tsstrasse, 25, D-33615,  Bielefeld,
Germany.  E-mail: {\it blanchard@physik.uni-bielefeld.de\/} }
\footnotetext[2]{ Fakult\"{a}t f\"{u}r Physik, Theoretische Physik,
Universit\"{a}t Bielefeld, Universit\"{a}tsstrasse, 25, D-33615,  Bielefeld,
Germany.  E-mail: {\it fortunat@physik.uni-bielefeld.de\/} }
\footnotetext[3]{PhyMat, D\'epartement de Math\'ematiques, UTV, BP 132, 
F-83957 La Garde Cedex, France.\\E-mail: {\it gandolfo@cpt.univ-mrs.fr\/} }
\footnotetext[4]{CPT, CNRS, Luminy case 907, F-13288 Marseille Cedex  9,
France.}

\thispagestyle{empty}
\footnotesize
\begin{quote}
{\sc Abstract:}
Recent results concerning the topological properties of random geometrical
sets have been
successfully applied to the study of the morphology of clusters in
percolation theory.
This approach provides an alternative way of inspecting the critical
behaviour of random systems in statistical mechanics.

For the 2d $q$-states Potts model with $q \le 6$, intensive and accurate
numerics indicates that the
average of the Euler characteristic (taken with respect to the
Fortuin-Kasteleyn random cluster
measure) is an order parameter of the phase transition.

\\[3pt]
%{\sc Pacs:} 05.20-y; 05.70Jk.
\end{quote}
\normalsize
\vskip15pt

%\baselineskip=14pt

% espace entre paragraphes :
\parskip=3pt plus 1pt minus 1pt

\section{Introduction}
\setcounter{equation}{0}

Recently, new insights in the study of the critical properties of clusters
in percolation theory have emerged based on ideas coming from mathematical
morphology  \cite{Se} and integral geometry \cite{Ha, S, Sch}. These
mathematical theories provide a set of geometrical and topological measures
allowing to quantify the morphological properties of random systems. In
particular these tools have been applied to the study of random cluster
configurations in percolation theory and statistical physics \cite{MW, O,
J, Wa1, M1}.

One of these measures is the Euler-Poincar\'{e} characteristic $\chi$ which
is a well known descriptor of the topological features of geometric
patterns. It belongs to the finite set of Minkowski functionals whose
origin lies in the mathematical study of convex bodies and integral
geometry (see \cite{Ha, S, Sch}).
These measures, as we shall explain below, share the following remarkable
property: any  homogeneous, additive, isometry-invariant and conditionally
continuous functional on a compact subset of the Euclidean space
$\mathbb{R}^{d}$ can be  expressed as a linear combination of the Minkowski
functionals. This is the well known Hadwiger's theorem  \cite{Ha} of
integral geometry which has a wide scope of applications in mathematical
physics due its rather general settings.

The use of these measures in image analysis \cite{Gra}, problems of shape
recognition \cite{Se}, determination of the large scale structures of the
universe \cite{Wa2}, modelling of  porous media \cite{AKPM}, microemulsions
\cite{HK} and fractal analysis \cite{M2} has been a topic of growing
interest recently.

The application of these description tools for the study of random systems
in statistical mechanics has already provided interesting results.
In \cite{MW,O},  the computation of the Euler-Poincar\'e characteristic
$\chi$ for a system of penetrable disks in several models of continuum
percolation has led to  conjecture new bounds for the critical value of the
continuum percolation density.
In \cite{MW}, an exact calculation of $\chi$ has shown that a close
relation exists between the zero of the Euler-Poincar\'e characteristic and
the critical threshold for continuum percolation in dimensions $2$ and $3$.

Of similar interest in the same domain, let us recall that for the problem
of bond percolation on regular lattices, Sykes and Essam \cite{SyEs} were
able to show, using standard planar duality arguments, that for the case of
self-dual matching lattices (e.g. $\mathbb{Z}^{2}$), the mean value of the
Euler-Poincar\'e characteristic changes sign at the critical point (this even
led them to announce a proof for the existence of the critical probability
of bond percolation on $\mathbb{Z}^{2}$), see also \cite{Gri}.

More recently, Wagner \cite{Wa1} was able to compute the Euler-Poincar\'e
characteristic on the set of all plane regular mosaics (the $11$
archimedean lattices) as a function of the site
occupancy probability $p\in[0,1]]$ and showed that a close connection exists
between the threshold for
site percolation on these lattices and the point where the Euler-Poincar\'e
characteristic (expressed
as a function of $p$) changes sign.

The aim of this article is to further investigate the role played by this
morphological indicator in
statistical physics and to present new results concerning its behaviour in
the case of  the
2-dimensional Potts model. Namely we present here clear evidence, based
on Monte Carlo simulations, that for
the 2d Potts model, the
Euler-Poincar\'e characteristic is an order parameter of the phase
transition. Namely we find
that for $q=2,...,4$ it changes sign continuously at the transition point
while, for $q=5, 6$ it has a
first order transition at the critical point. As far as we know, this is
the first example of a
discontinuous behaviour of this parameter in a physical model.

The paper is organized as follows. In Section 2 we introduce a minimal
background concerning Minkowski functionals and the necessary definitions
for our model, the numerical results are presented in Section 3 followed by
some
comments and discussion in Section 4.

\section{The model}
\setcounter{equation}{0}

We first briefly summarize the basic facts from integral geometry and give
the definition of the Minkowski functionals including the Euler-Poincar\'e
characteristic, see \cite{Ha, S, Sch} for more complete expositions. We
will then show how to compute the Euler-Poincar\'e characteristic in the
case of a random configuration of sites and bonds produced by
a Fortuin-Kasteleyn (FK) transformation of the partition function of the
Potts model \cite{FK}.

\subsubsection*{Minkowski Functionals}
\setcounter{equation}{0}
In the most abstract setting, the Minkowski functionals are defined as
integrals of curvatures in the framework of the differential geometry of
smooth surfaces in compact sub--domains of the Euclidean space $\Reel^d$
\cite{Ha}.

Another way to introduce these measures is as follows.
 First define the Euler-Poincar\'e
characteristic $\chi$ as an additive functional on subsets of $\Reel^d$
such that for $A, B\subset \Reel^d$ (see example in Fig. 1)
\begin{equation}
\chi(A\cup B) = \chi(A) +  \chi(B) - \chi(A\cap B)
\label{EC}
\end{equation}
with
$$
\chi(A) =
\begin{cases}
1, \; \mbox{if} \; A \ne \emptyset \mbox{,  A convex} \\
0, \; A=\emptyset.
\end{cases}
$$

\begin{figure}[htb]
\begin{center}
\epsfig{file=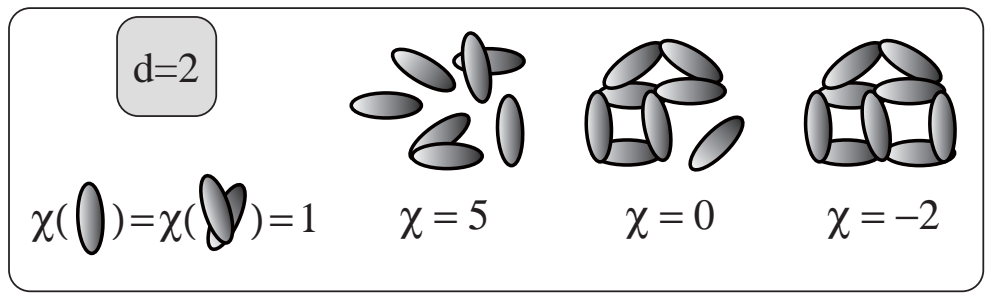,width=13cm}
\caption{\label{fig1}{Examples of calculation of the Euler-Poincar\'e characteristic in
dimension $d=2$ for some combinations of convex subsets of $\Reel^2$.}}
\end{center}
\end{figure}

The set of Minkowski functionals is then defined by (see e.g. \cite{Gra,MW})
\begin{eqnarray}
W_{\a}(A) &=& \int \chi(A\cup E_{\a}) \, d\mu (E_{a}), \;\;\; \a=0,...,d-1 \\
W_{d}(A)  &=& \o_{d}\, \chi(A) : \;\;\; \o_{d}=\pi^{d/2} / \Gamma(1+d/2)
\end{eqnarray}
where $E_{\a}$ is an $\a$-dimensional plane in $\Reel^d$, $d\mu (E_{a})$
its density normalized such
that, for the $d$-dimensional ball $B_{d}(r)$ with radius $r$,
$W_{\a}(B_{d}(r))=\o_{d} \, r^{d-\a}$ and
$\o_{d}$ is the volume of the unit ball in $\Reel^d$.

Obviously, additivity of the Minkowski functionals is inherited from
(\ref{EC}), furthermore they are
conveniently  normalized through
$$
M_{\a}(A) = \frac{\o_{d-\a}}{\o_{d} \, \o_{\a}} W_{\a}(A)
$$
The computation of these normalized functionals in dimensions $1,2,3$ in
term of the usual geometric measures (length, area, volume,...) is given in
Table 1.

\vskip0.5cm
\begin{table}[ht]
\begin{center}
\begin{tabular}{ p{0.5cm} || c | c | c | c }
$d$     &       $\;\;\;M_0\;\;\;$       &       $\;\;\;M_1\;\;\;$       &
        $M_2$   & $M_3$ \\ \hline \hline
1       &       $L$     &       $\chi_{1}/2$    &       $\cdots$        &
        $\cdots$    \\ \hline
2       &       $S$     &       ${\cal C}$      &       $\chi_{2}/\pi$  &
        $\cdots$    \\ \hline
3       &       $V$     &       $S$             &       $H/2\pi^{2}$    &
        $3\chi_{3}/4\pi$ \\
\end{tabular}
\caption{Values of the normalized Minkowski functionals for $d$-dimensional
subsets of $\Reel^d, d=1,2,3$ in terms of geometric measures, $L$: length, $S$: area, $V$: volume, ${\cal C}$: circumference, $H$: integral mean curvature, $\chi_{d}$: Euler-Poincar\'e characteristic.}
\end{center}
\end{table}

Hadwiger's completeness theorem asserts that, under not too restrictive and
furthermore physically reasonable assumptions, namely: {\it additivity},
{\it motion invariance} (under translations and rotations) and {\it
conditional continuity} (which states that any convex body can be smoothly
approximated by convex polyhedra), any functional ${\cal M}$ decomposes as
a linear combination of the {\it finite} set of Minkowski functionals
$$
{\cal M(A)} = \sum_{\nu=0}^{d} c_{\nu} \, M_{\nu}(A)
$$
where the $c_{\nu}$ are real coefficients.

This theorem has very important practical consequences, we refer to
\cite{Ha,S,Gra}, we shall only mention the {\it principal kinematic
formula} (see \cite{S, M1})
$$
\int_{\cal G} M_{\a}(A\cup g B) dg = \sum_{\b=0}^{\a} {\binom \a \b}
M_{\a-\b}(B) M_{\b}(A), \; \a=0,...,d
$$
where integration is over the group of motions (i.e. rotations and
translations, $g=(r,\Theta)$).
This formula is very useful to calculate mean values of Minkowski
functionals for random distributions of objects. For instance, it can be
applied to the computation of the excluded volume of convex bodies
(leading,  in case of spherical objects, to Steiner's formula), which has
been used to estimate the critical thresholds in continuum percolation
theory \cite{Ba, PS}.

In the following, we will be mainly interested in the evaluation of the
Euler-Poincar\'e characteristic for random bond configurations on regular
lattices $\L \subset \rel^d$ which, as we shall see, is equivalent to
compute Euler formula for planar graphs made of ${\cal S}$ sites (or nodes), ${\cal B}$
bonds (or links) and ${\cal P}$ plaquettes (or faces) \cite{Ag, Be}, i.e. 
$$
\chi={\cal S} - {\cal B} + {\cal P}
$$
It is indeed one of the topics of algebraic topology to show that the above
general definitions of Minkowski functionals extend to cell complexes (see Appendix) 
to which random bond configurations on regular lattices belong.

Because of the growing interest in the use of these morphological measures,
we thought that it was important to recall briefly their definitions and
main properties.

\subsubsection*{Potts Model}

The  partition function for the $q$-states Potts model on $\Lambda \subset
\mathbb{Z}^{d}$
at inverse temperature $\beta$ reads
\begin{equation}
Z_{\beta,q}^{\rm Potts}(\Lambda)=
\sum_{\bs}
\exp\bigg\{  \beta \sum_{<i,j>\subset \Lambda}
\delta(\sigma_{i},\sigma_{j})\bigg\}
\label{PF}
\end{equation}
where the first sum runs over all configurations $\bs \subset \{1,\dots,q\}^{|\Lambda|}$, 
the second one is over each nearest neighbour pair of Potts spins on $\Lambda$ and $\delta$ is the Kronecker symbol.
We remind that, whenever $q$ is large enough, in any dimension $d\ge 2$,
this model exhibits
a unique (inverse) temperature $\beta_{c}$ where the mean energy is
discontinuous (see \cite{K,L1,L2}).
In dimension $d=2$ for $q\ge 5$ this is an exact result \cite{Bax} and it
is expected to be true, in $d=3$, for $q \ge 3$ \cite{Wu}.

After performing the (FK) transformation \cite{FK},
the partition function (\ref{PF}) leads  to the following  random cluster
representation
\begin{equation}
Z_{\beta ,q}^{\rm FK}(\Lambda) = \sum_{X} (e^{\b}-1)^{N^1(X)} q^{N_{\L}(X)}
\label{RCR}
\end{equation}
Here the summation is over all graphs $X$ which can be drawn inside the
domain $\Lambda$,
$N^1(X)$ is the number of bonds of the configuration $X$
and $N_{\L}(X)$ is the number of connected components of  $X$ (including
isolated sites).
We will call
$N^0(X)$ the number of sites which are endpoints of a bond in $X$ and
$N^2(X)$ the number of
plaquettes of the configuration $X$, i.e. the set of cells in $\L$ having 4
occupied bonds on its boundary (see Fig. 2).

\vskip0.5cm
\begin{figure}[htb]
\begin{center}
\epsfig{file=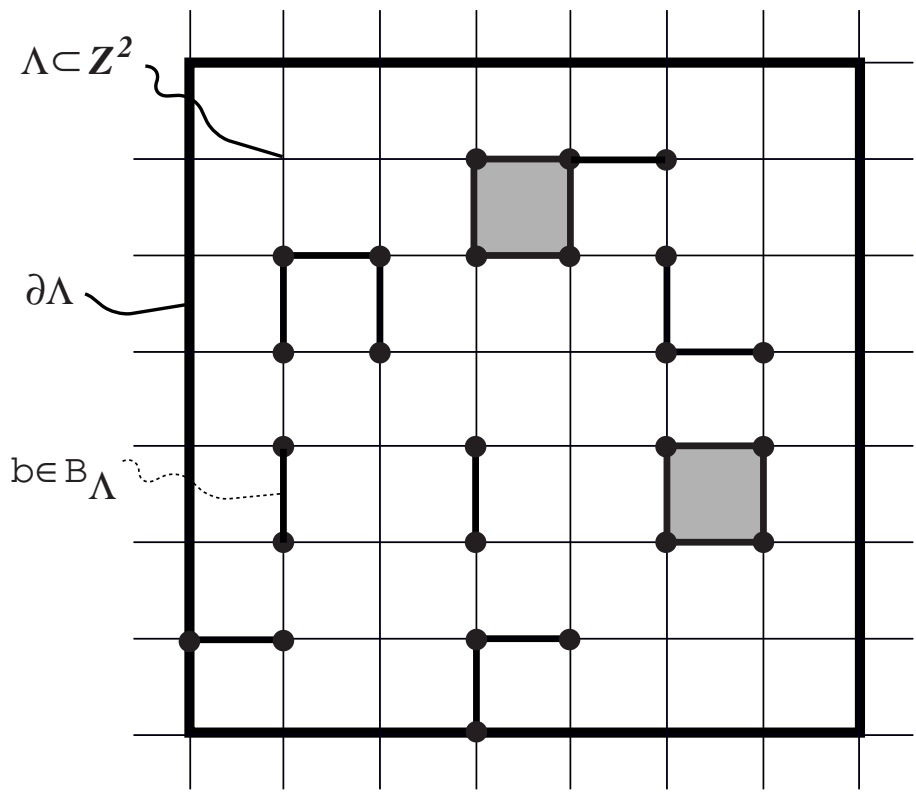, width=12cm}
\caption{A bond configuration $X\subset B_{\Lambda}$ with 25 sites, 19
bonds and 2 plaquettes.}
\end{center}
\end{figure}

In this framework, the Euler-Poincar\'e formula \cite{Ag} leads to the
following definition of the Euler-Poincar\'e characteristic
\begin{equation}
\chi(X) = N^{0}(X) - N^{1}(X) + N^{2}(X)
\label{ECconf}
\end{equation}
This expression will allow us to compute the mean value of the Euler-Poincar\'e
characteristic with respect to the FK measure. 

In algebraic topology, one is interested in classifying geometric objects (or spaces) up to some transformations acting on these spaces (homeomorphisms) by means of simple invariant quantities. It turns out that this program can be achieved and such invariants of a space are the homology groups of these spaces. In this framework, the Euler-Poincar\'e characteristic (as the other Minkowski functionals) can be expressed in terms of these invariants and lead to a definition for $\chi$ equivalent to the one given by (\ref{ECconf}) (see Appendix).

\section{Numerical Results}
\setcounter{equation}{0}

We have performed Monte Carlo simulations of the 2-dimensional
$q$-state Potts model for $q$ ranging from $2$ to $6$.
We have always simulated the models near  the critical
temperature, whose value can
be exactly determined through the well known
formula: $\beta_c(q)=J/kT_c(q)=\log(1+\sqrt{q})$.
In order to extract a value of the Euler characteristic
$\chi$ as close as possible to the value at the infinite volume
limit, we have taken rather large lattices: for the three
models with a continuous transition ($q=2,3,4$) we arrived at lattice sizes
up to $2000^2$. The algorithm we used is the Wolff cluster update
\cite{Wo}. The identification
of FK cluster configurations has been performed via the
 Hoshen--Kopelman algorithm \cite{Ho}; we always
considered free boundary conditions for the cluster labeling.

The Wolff algorithm is the less efficient the bigger the number
$q$ of states.
Particularly dramatic is what happens when one passes from the
2-state (Ising) to the 3-state model: in the former case, on the
$2000^2$ lattice it is enough to perform few updates ($5-10$)
to get uncorrelated configurations for the cluster variables,
in the latter one needs about 1000 updates!
Because of that, the simulations for $q=3,4$ on the $2000^2$
lattice were very slow, and the relative
data could not reach a high statistics. Nevertheless, as we will see,
we can get useful indications out of them.

If $\chi$ changes sign at the threshold,
on a finite lattice the values measured at
each iteration would be distributed
around zero, provided the lattice
is large enough. Therefore one would see both positive and negative values.
For this reason, it is helpful to look at the distribution of $\chi$.
In the following we present separately the results
for $q=2,3,4$ and $q=5,6$.

\subsection{Results for the models with a continuous phase transition}
\setcounter{equation}{0}

The first case we consider here is the Ising model. Fig. 3
shows the $\chi$ distribution for three different lattice sizes:
$500^2$, $1000^2$ and $2000^2$. In each case we have
taken 20000 measurements.
The peak of the distribution
shifts towards $\chi=0$ the larger the lattice. The average values of $\chi$
are: $\chi(500^2)=0.00101(2)$, $\chi(1000^2)=0.00053(2)$,
$\chi(2000^2)=0.00024(1)$.
We notice that the averages are quite small and
decrease sensibly if we go to larger
sizes, reducing themselves to about the half when we pass from a lattice
to the next one. This approximate linear scaling of $\chi$ with the lattice
side suggests that the Euler characteristic at the infinite volume
limit indeed vanishes.

Let us now examine the case $q=3$. In Fig. 4 we
again plot the $\chi$ distribution for the same three lattice sizes
we have considered for the Ising model.

\begin{figure}[htb]
\begin{center}
\epsfig{file=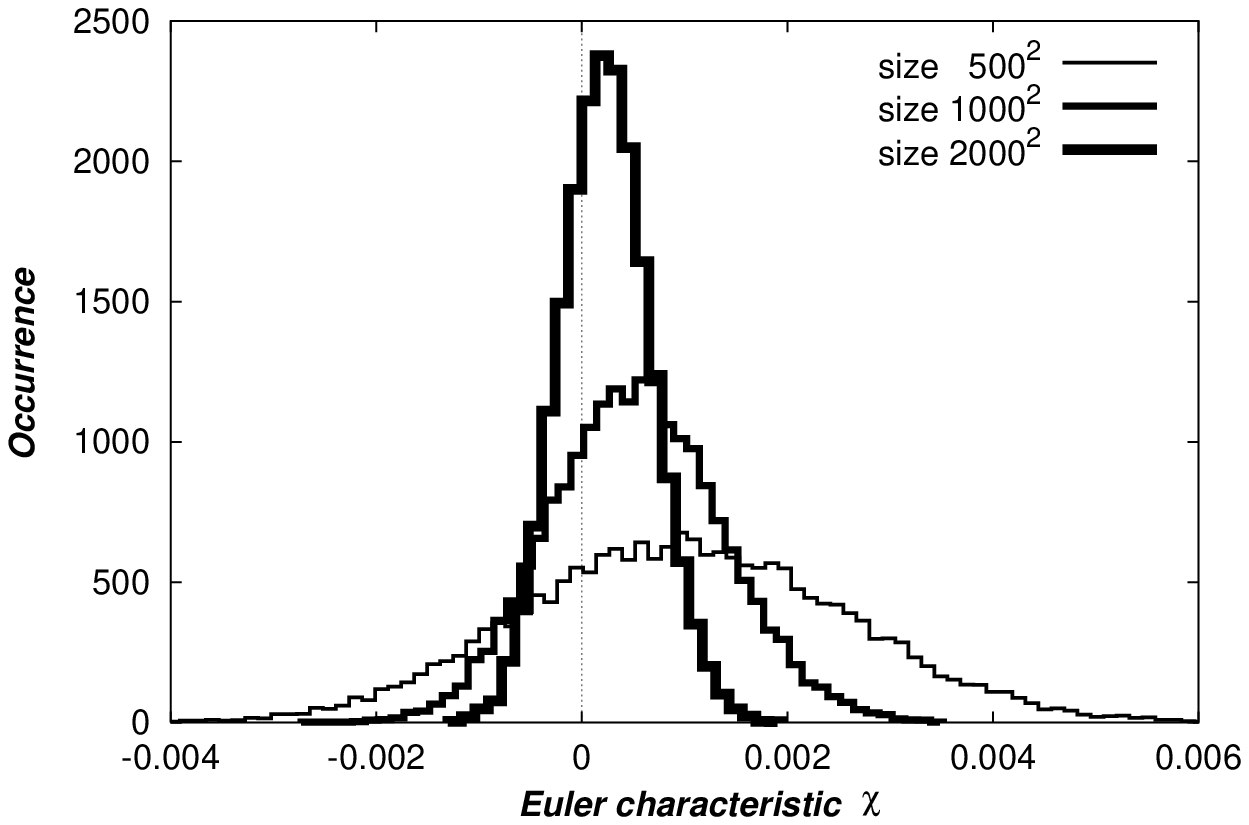, width=15cm}
\caption{Distribution of $\chi$ for the FK cluster
configurations of the 2d Ising model at the critical point.}
\epsfig{file=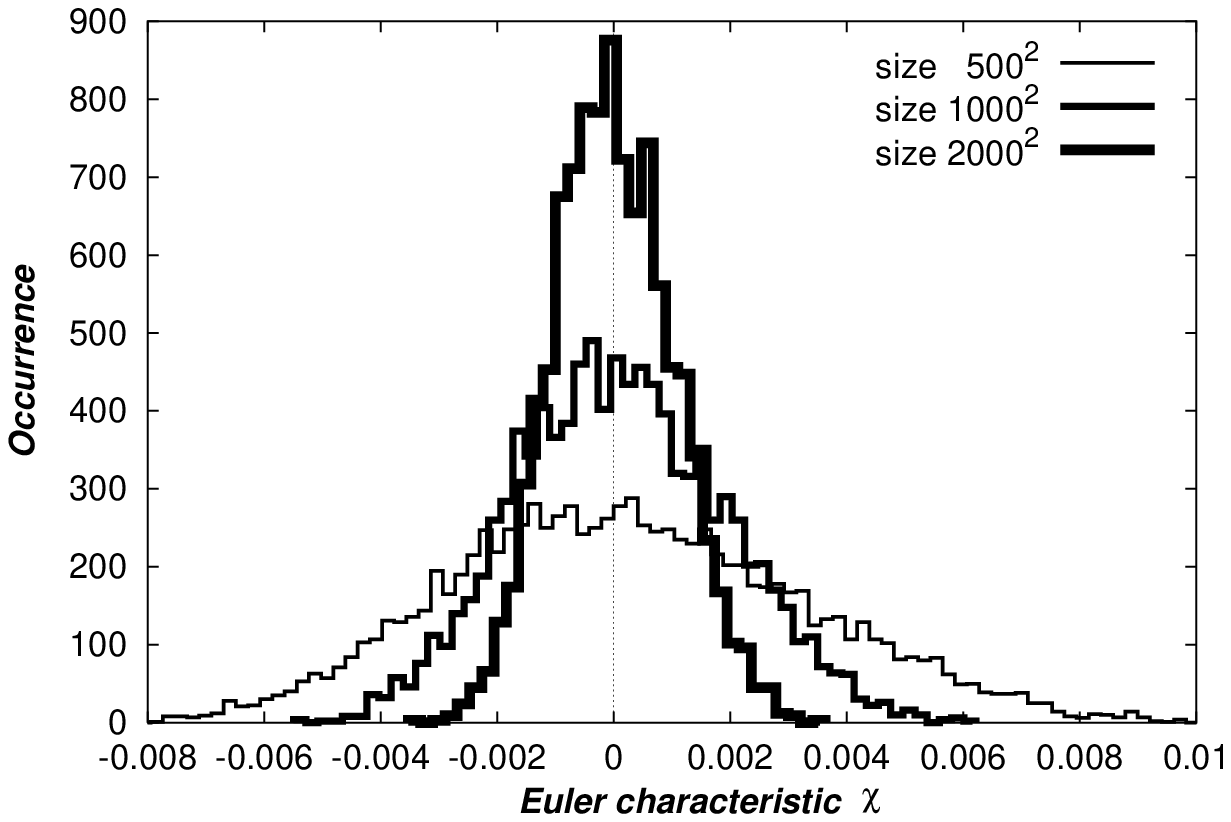, width=15cm}
\caption{Distribution of $\chi$ for the 
FK  configurations of the 2d, 3-state Potts model at the critical point.}
\end{center}
\end{figure}

\clearpage

Since we collected a different number of measurements
for the different lattices, for a real comparison 
of the distributions
we needed to renormalize the total number of measurements on each
lattice to the same
value: we decided to renormalize
all data sets to the number of measurements on the
$500^2$ lattice (10000).
The distributions are broader than
the Ising ones but they
appear almost exactly centered at $\chi=0$. 
The average values
are in fact much smaller than before:
$\chi(500^2)=0.00024(4)$, $\chi(1000^2)= 0.00012(3)$,
$\chi(2000^2)$ is zero within errors. We then deduce that also
for $q=3$ $\chi=0$ at the critical point.

To complete our analysis we studied the case $q=4$. In Fig. 5
we present a comparison of the $\chi$ distributions for
two lattice sizes, $800^2$ and $2000^2$.

\begin{figure}[htb]
\begin{center}
\epsfig{file=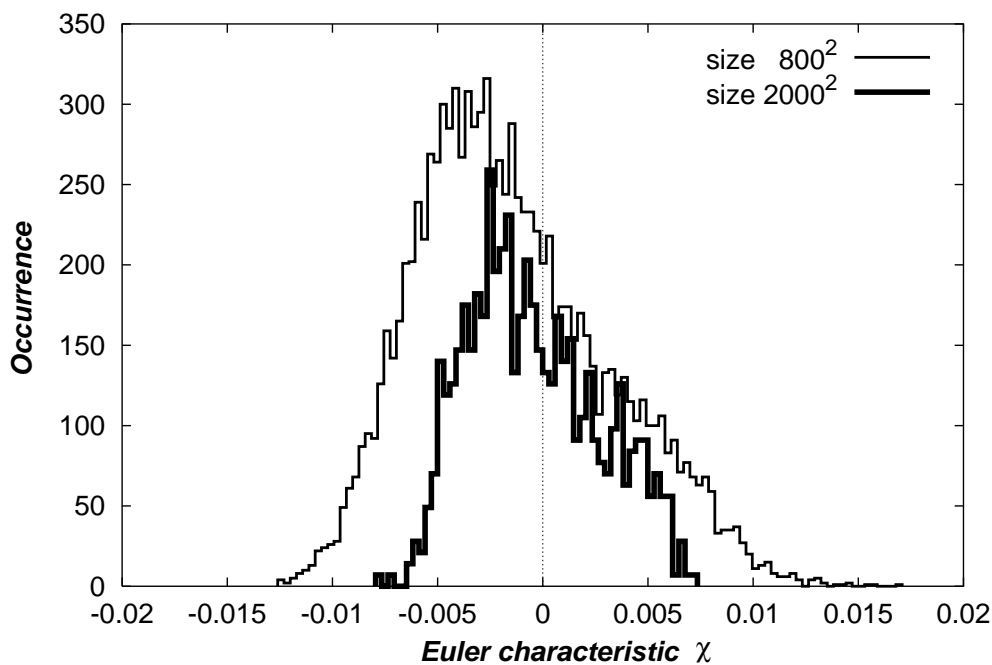, width=15cm}
\caption{Distribution of $\chi$ for the FK cluster
configurations of the 2d, 4-state Potts model at the critical point.}
\end{center}
\end{figure}

There is a clear shift of the center
of the distribution towards zero when one goes from the smaller
to the larger lattice. The average values of the Euler characteristic
in the two cases are
$\chi(800^2)=-0.00145(8)$ and $\chi(2000^2)=-0.0006(2)$. Also here there is
no apparent convergence to some value, even if the lattices are rather large:
$|<\chi>|$ reduces itself to less than its half by changing the lattice size.
From all this we also deduce that the Euler characteristic
of the FK clusters of the 2-dimensional 4-state Potts model vanishes
at criticality.

\newpage

\subsection{Results for the models with a first order phase transition}
\setcounter{equation}{0}

For $q>4$ the 2d $q$-state Potts model undergoes a first order phase transition, i.e.
the thermal variables vary discontinuously at the critical threshold.
The magnetization, for instance, makes a jump,
varying from zero to a non-zero value.
Because of that, we expect that the cluster configurations change
abruptly at the critical point, and that the cluster variables
exhibit as well discontinuities. In particular, the Euler characteristic
may jump from a value to another.

We analyze here the 5- and 6-state Potts models. In both cases we have
performed simulations on a $300^2$ lattice. In Figs. 6 and 7 we
compare the distribution histograms of the magnetization $M$ and the 
Euler characteristic $\chi$ at three different temperatures:
above, near and below $T_c$. We define the magnetization by taking the excess
of sites in the majority spin state with respect to the value
$1/q$ in the paramagnetic phase, when all spin states are equally distributed. 
Therefore we always measure $M>0$. Looking at the 
magnetization histograms one clearly sees  
the spontaneous symmetry breaking by reducing the temperature.
The characteristic double peak structure of $M$ around $T_c$
indicates that the transition is first order, as it is known. 
The corresponding histograms of the Euler characteristic show a perfectly analogous 
pattern. As we expected, the two coexisting FK-phases at $T_c$
are characterized by two different values of the Euler characteristic
(double peak in the figures). The ordered phase is made of bonds between Potts spins of the same color at $T_c$. We then deduce that
$\chi$, like $M$, varies discontinuously at the threshold.
The result is valid for the 5-state and the 6-state Potts model,
so it is likely to be valid also for $q>6$, when the discontinuity of $M$
at the threshold is sharper.
Looking at both figures we remark that the centers of the peaks 
of $\chi$ look approximately symmetric with respect to zero. If this symmetry exists,
it would be an interesting feature, and at the moment we have
no arguments to justify it. 
In order to determine with some accuracy the values of $\chi$ in the two 
coexisting phases we would need
to increase considerably the size of the lattice, 
but the required computer time 
would increase dramatically for the reasons 
we explained at the beginning of this section.

\begin{figure}[htb]
\begin{center}
\epsfig{file=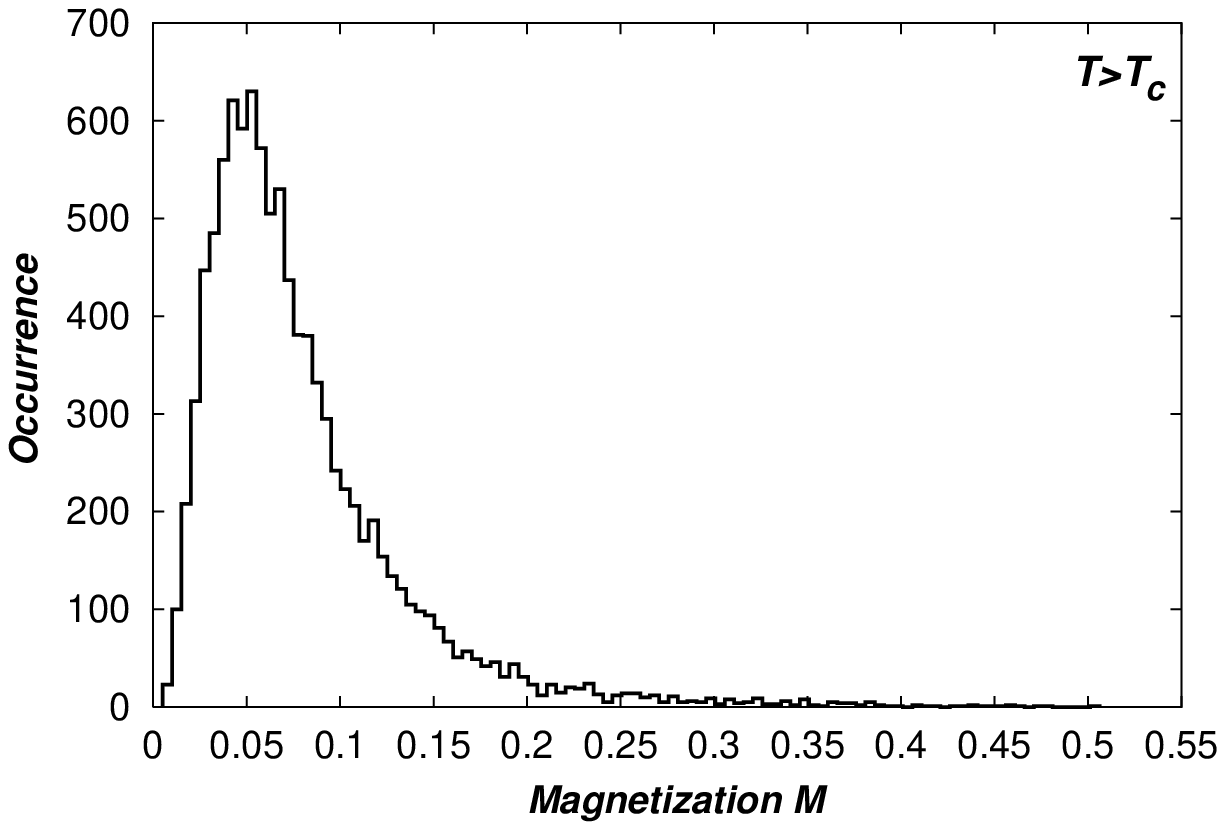,width=7.4cm}
\epsfig{file=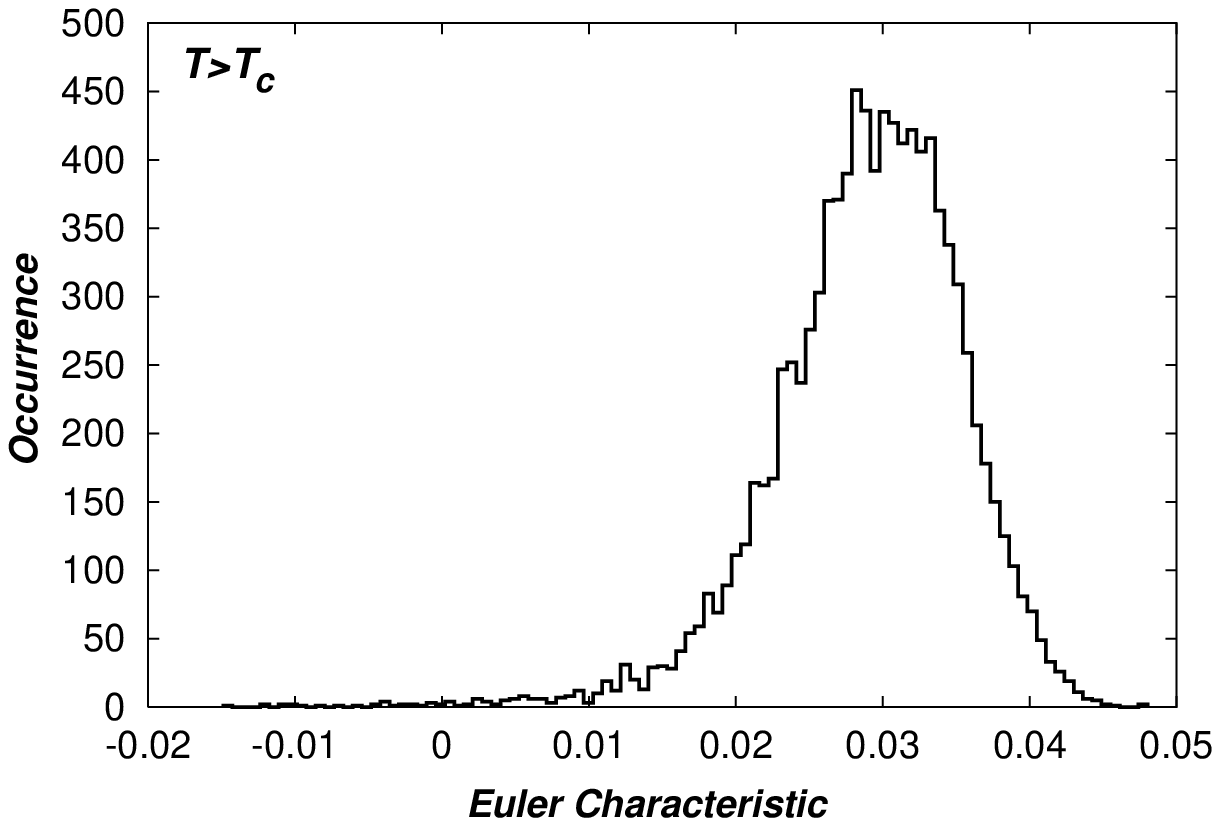,width=7.4cm}
\epsfig{file=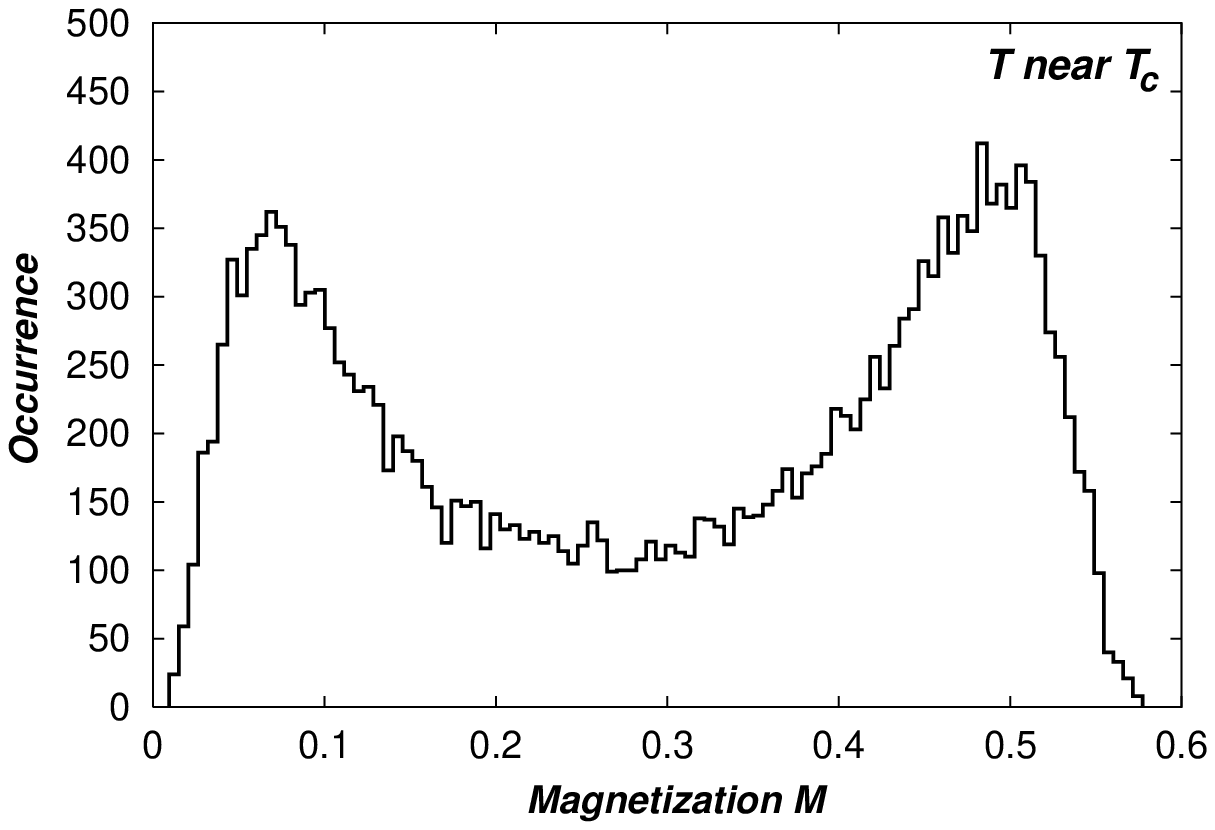,width=7.4cm}
\epsfig{file=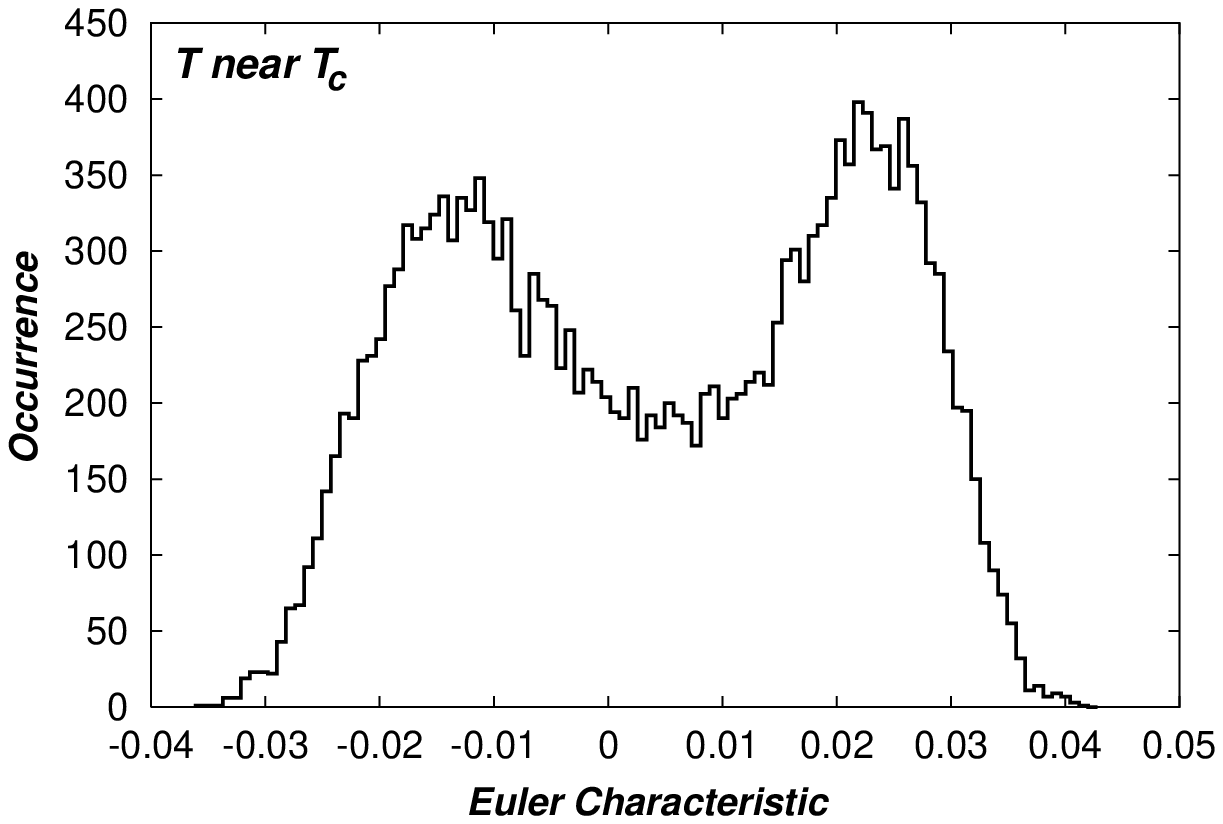,width=7.4cm}
\epsfig{file=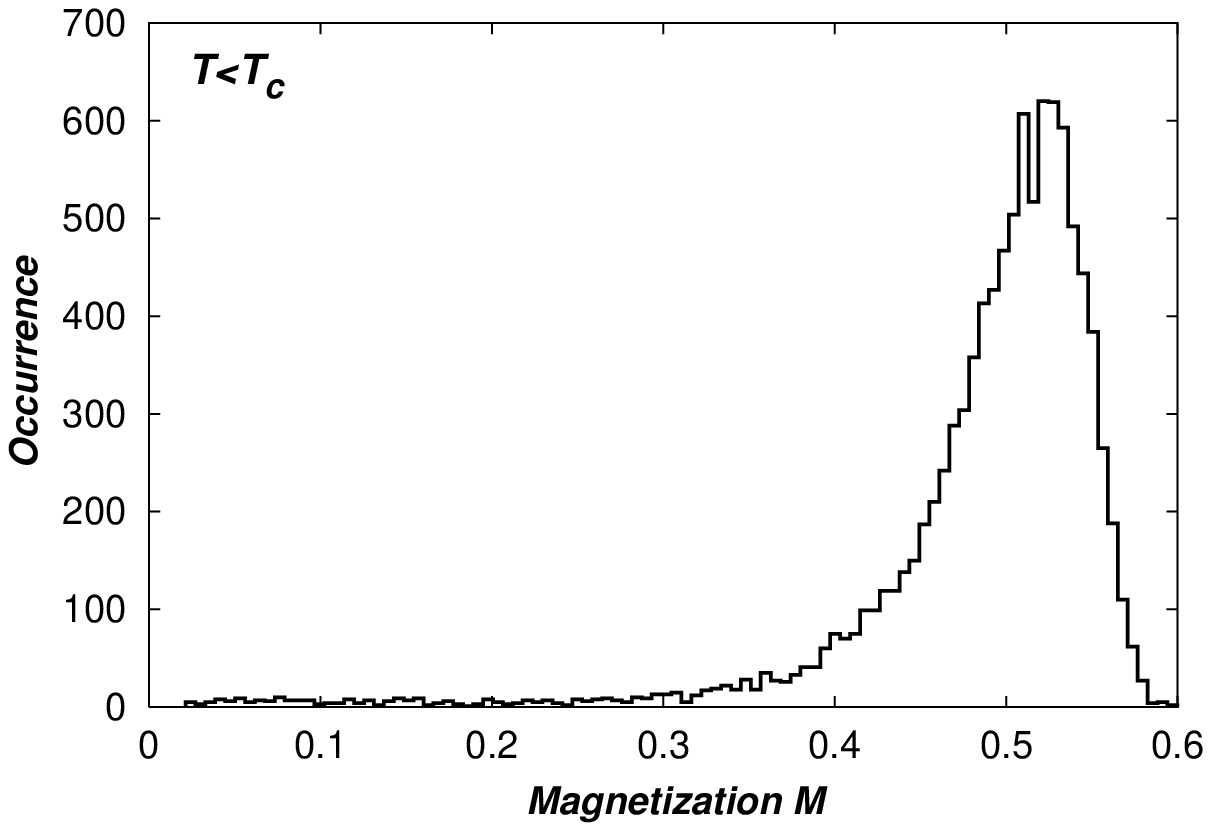,width=7.4cm}
\epsfig{file=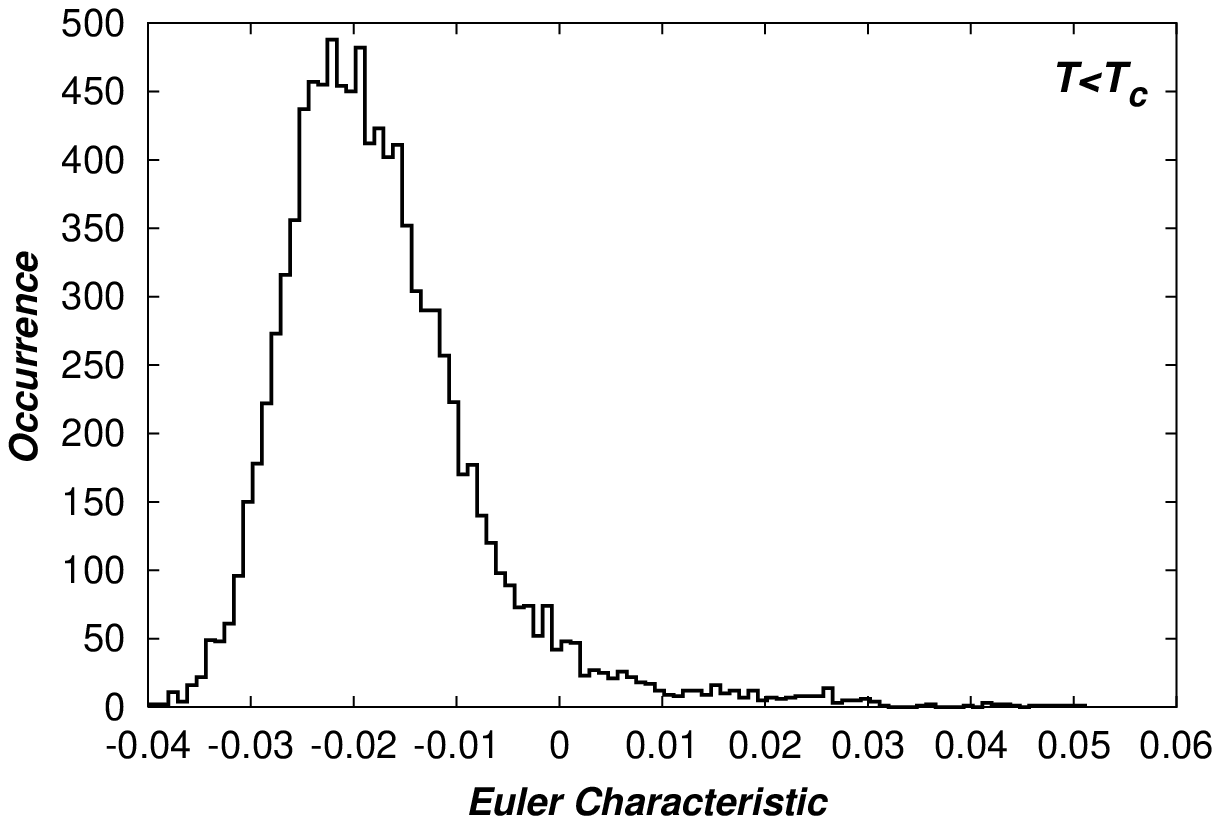,width=7.4cm}
\caption{Distribution histograms of the magnetization $M$ and the Euler characteristic $\chi$ for the 2-dimensional 5-state Potts model at three different temperatures. The lattice size is $300^2$. The behaviour of $\chi$ is driven by $M$.}
\end{center}
\end{figure}
\clearpage

\begin{figure}[htb]
\begin{center}
\epsfig{file=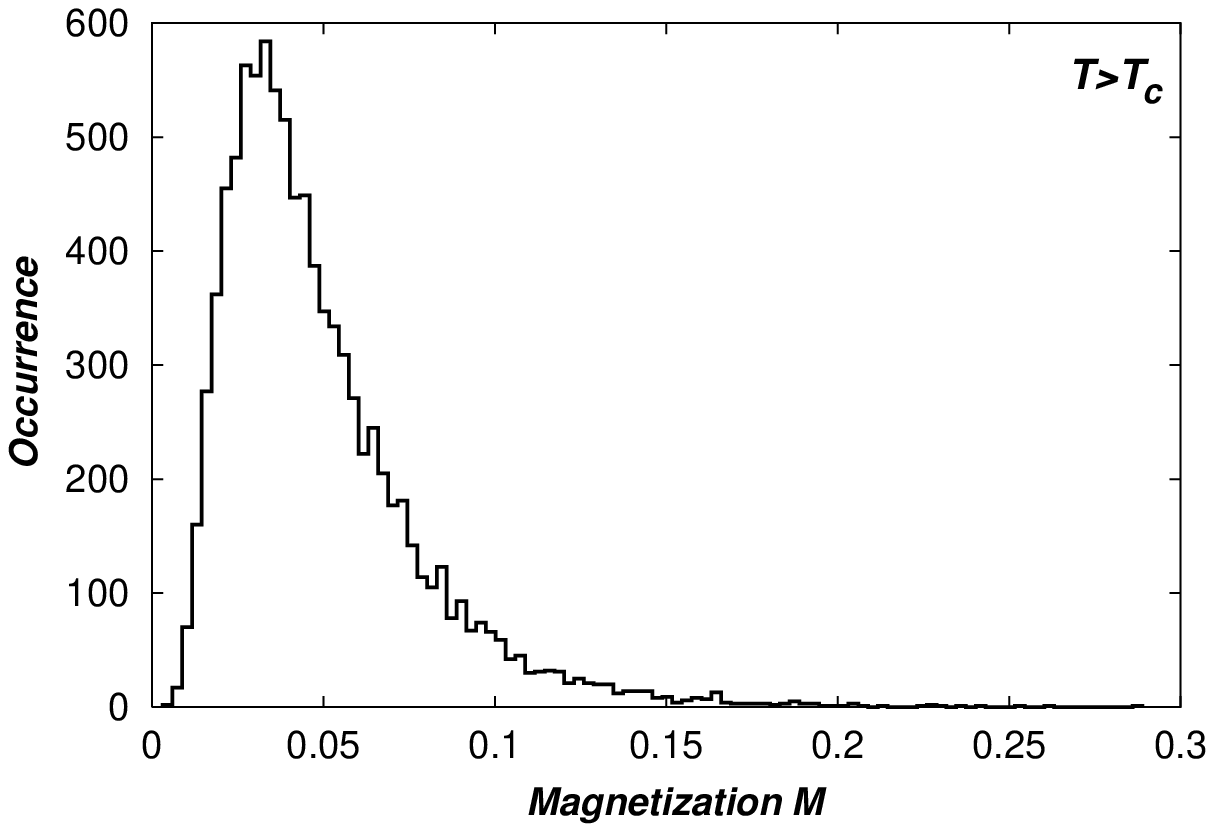,width=7.4cm}
\epsfig{file=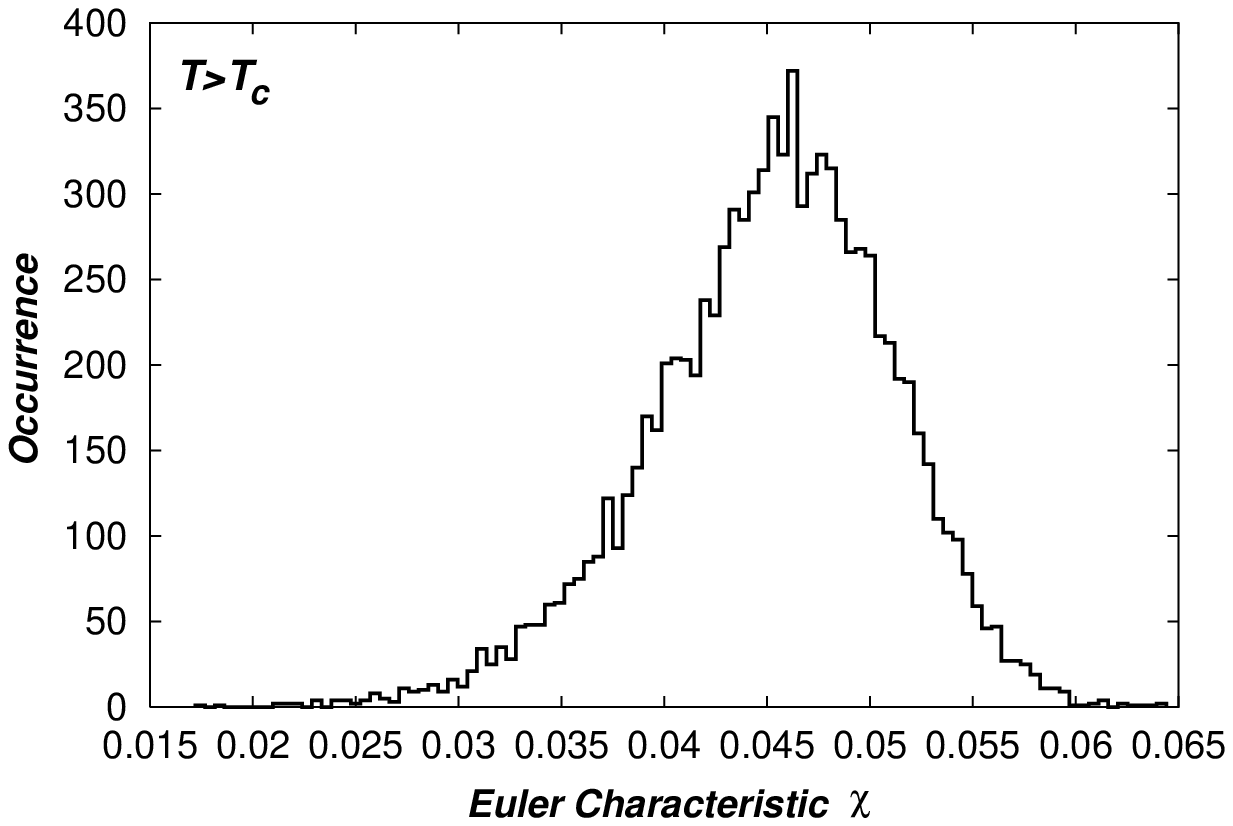,width=7.4cm}
\epsfig{file=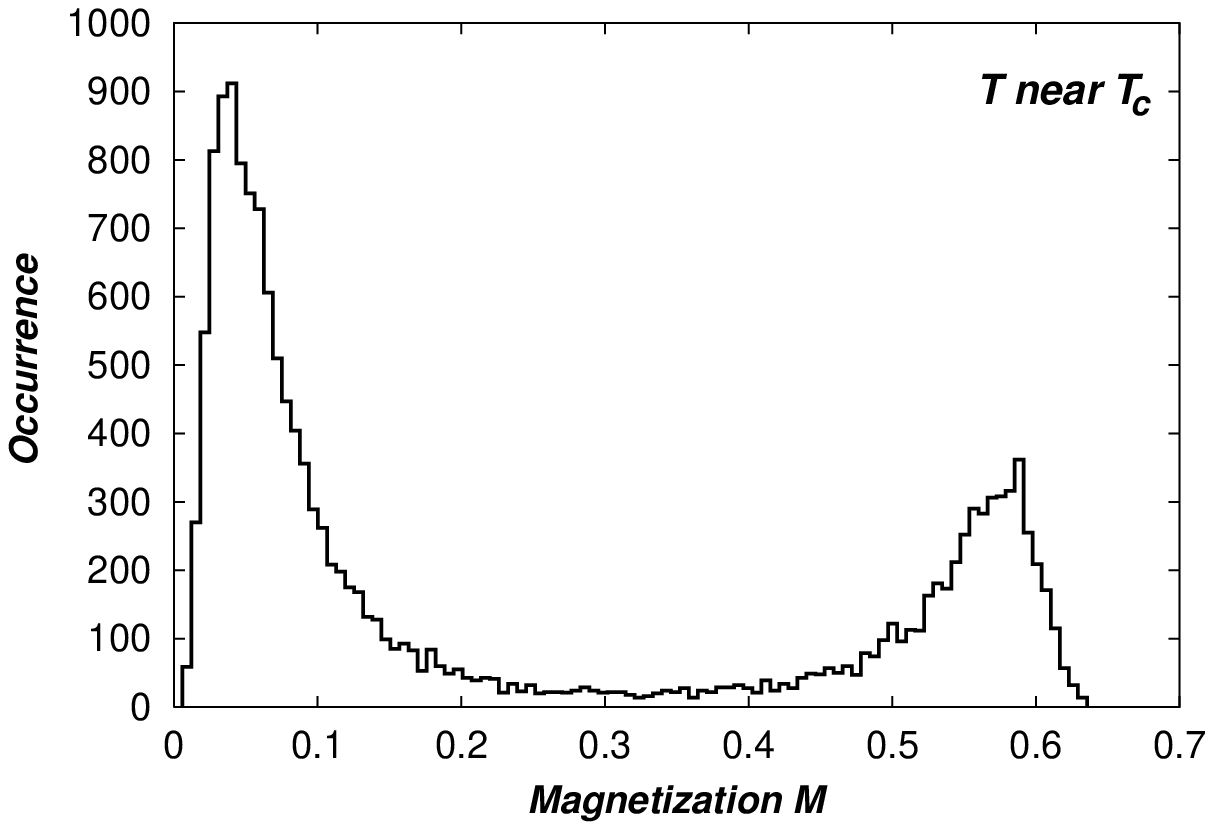,width=7.4cm}
\epsfig{file=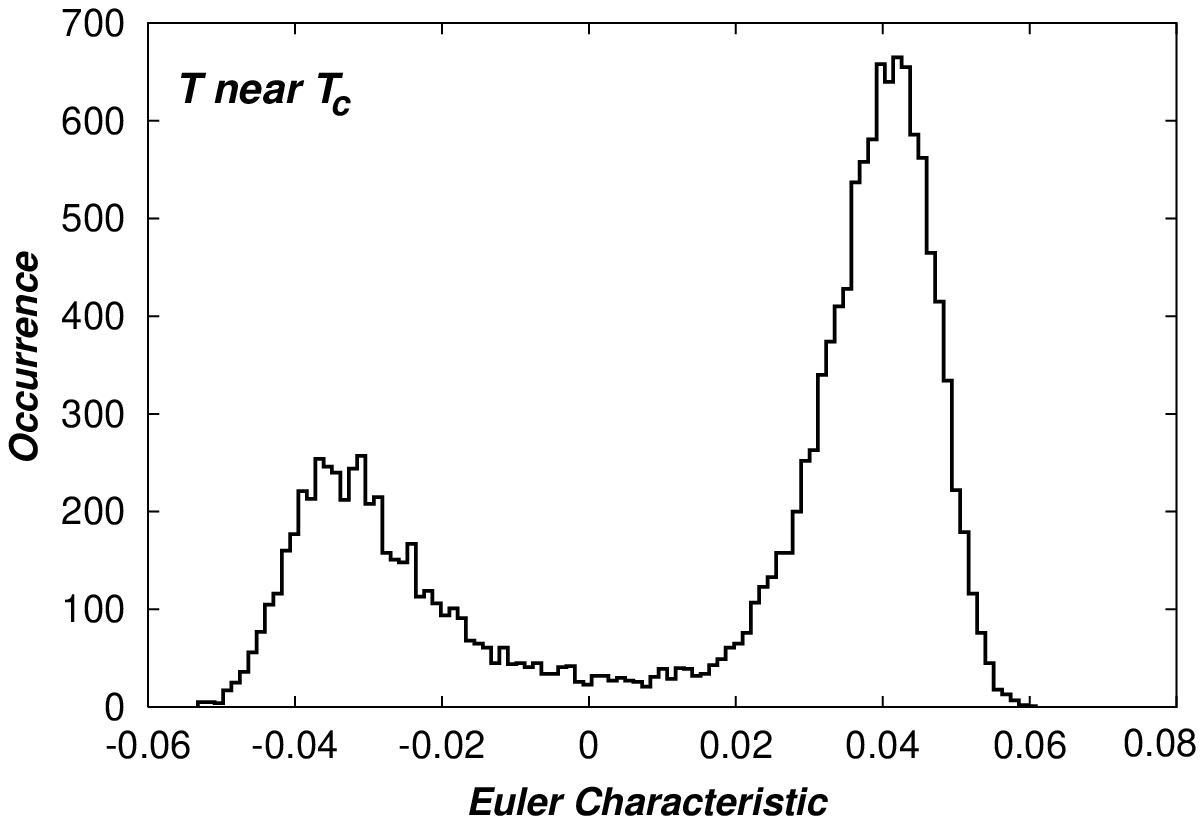,width=7.4cm}
\epsfig{file=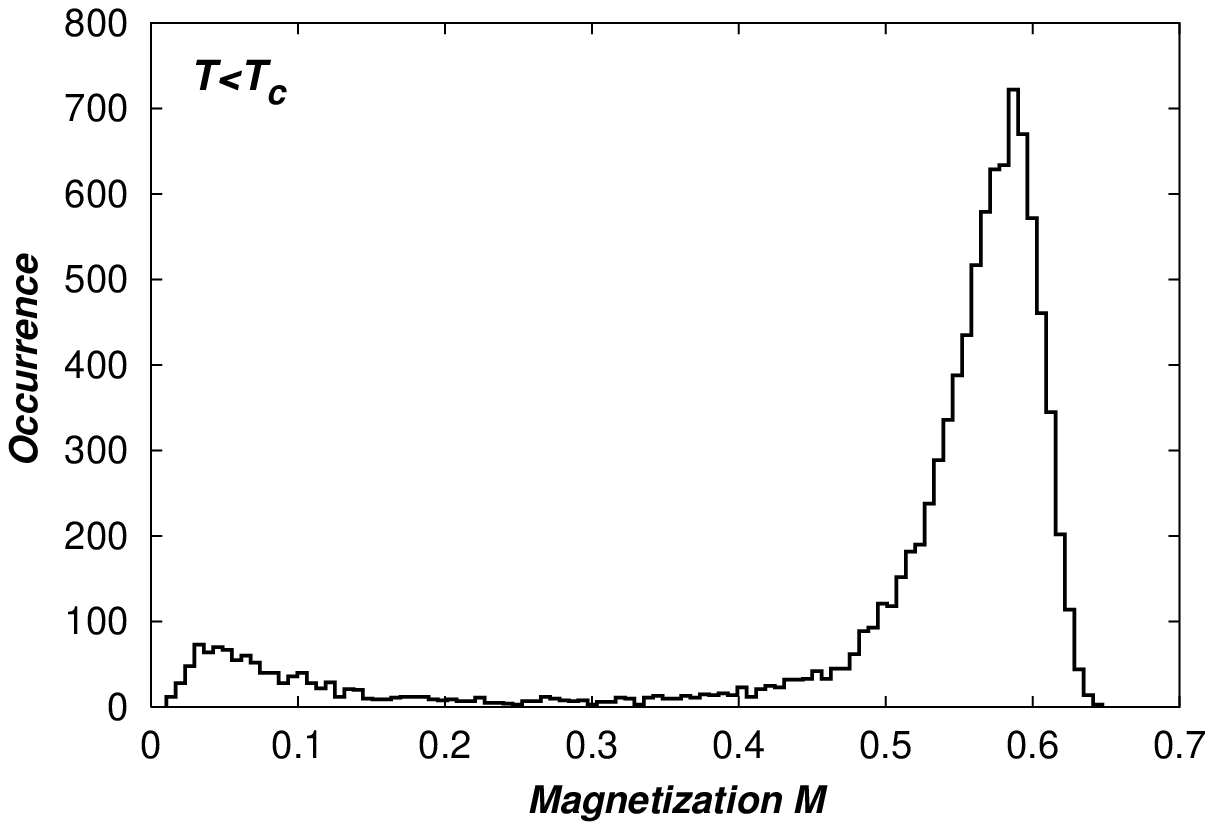,width=7.4cm}
\epsfig{file=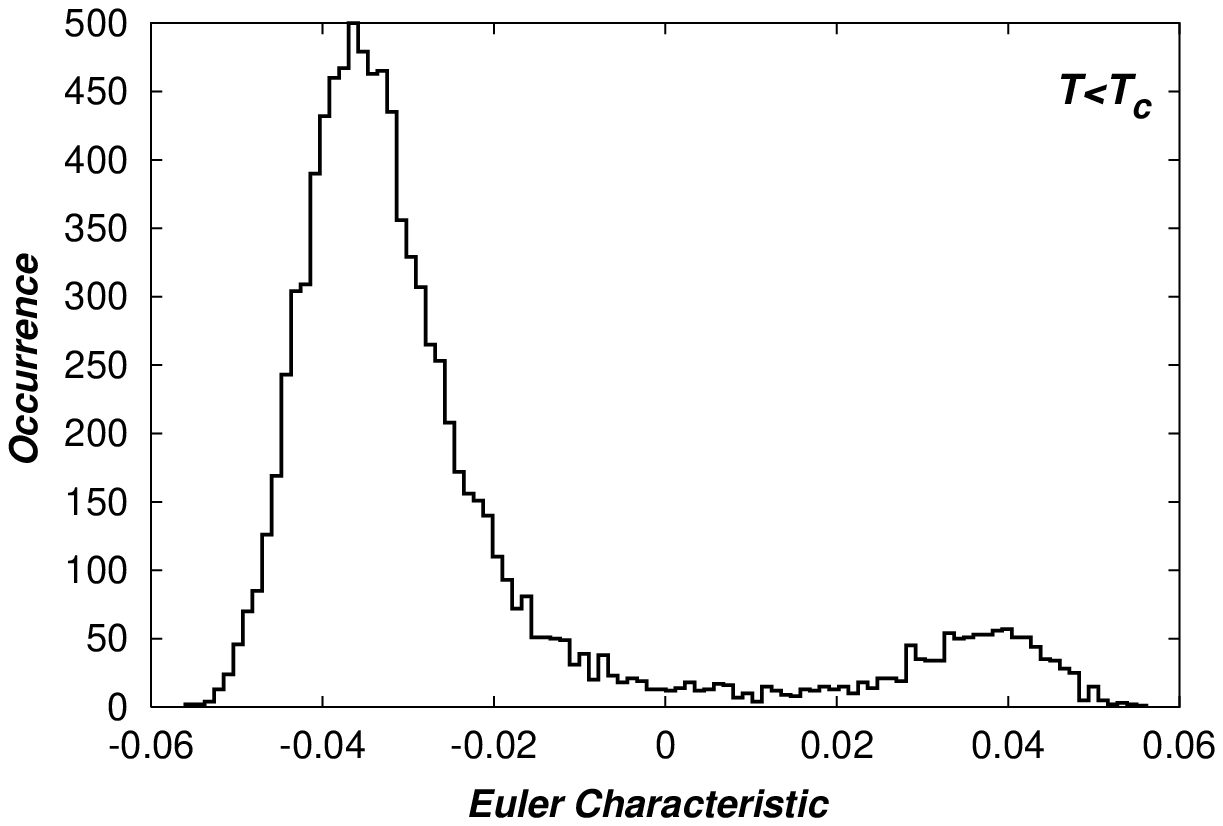,width=7.4cm}
\caption{Distribution histograms of the magnetization $M$ and the Euler characteristic $\chi$ for the 2-dimensional 6-state Potts model at three different temperatures. The lattice size is $300^2$. The behaviour of $\chi$ is driven by $M$.}
\end{center}
\end{figure}
\clearpage
\newpage
\section{Conclusions}
\setcounter{equation}{0}
This work clearly indicates and confirms that the Euler-Poincar\'e
characteristic is indeed an important indicator of a phase transition,
playing the role of an order parameter for the $2$-$d$ Potts model, to the
extend of the cases ($q=2,...,6$) studied here. For this model, it reveals
that the topology of cluster configurations has a deep meaning concerning
criticality.

The fact that $\chi$ changes sign at $T_c$ can be understood in the following way.
Let us consider the 2D Ising model: from $T_c$ up to $T=\infty$ the system is in its disordered phase and the only excitations one can get in the FK-bond representation are made of
isolated bonds (the probability to see any plaquette vanishes exponentially). Applying (\ref{Betti}) (see Appendix), one sees that $\chi$ behaves like $\b^{0}(X)$ times a term of the order of the volume of the system. 
However, from $T_c$ down to $T=0$, the system is in its ordered phase and the corresponding FK-configuration is (with high probability) made of $O(1)$ connected bond components. Missing bonds constitute the excitations and their number scales with the volume of the system so, using again (\ref{Betti}), one gets that $\chi$ behaves like $-\b^{1}(X)$ times a term of the order of the volume of the system. This explains in the case of the Ising model the change of sign of $\chi$ at $T_c$.

Other spin systems have to be investigated in order  to see
whether this property of the Euler-Poincar\'e characteristic is shared 
by models with continuous symmetries such as the
$X$-$Y$-model or the Widom-Rowlinson model.

Another important question concerns the critical behaviour in gauge models.
For example a similar study could  provide some
insights concerning the deconfining transition in SU(N) gauge
theory. Indeed, some works \cite{Sa} tend to indicate that this transition
could be probed by percolation of some physical clusters related to color
fields in lattice QCD. These models have been thoroughly
investigated in the past and the tools coming from algebraic topology have
been of primary importance to uncover profound duality results concerning
their phase structure \cite{We, Wa3}.

The interest of the last remark resides in the fact that so far, no
suitable order parameter is known that can  describe the deconfining
transition in SU(N) gauge
theory in the case of finite quark masses and it would be interesting to
figure out how does the Euler-Poincar\'e
characteristic behave in gauge models.

\section{Acknowledgments}
\setcounter{equation}{0}

We are indebted to H. Wagner who initiated our interests for the topics
developed in this paper and to J. Ruiz for fruitful discussions. 

Financial support from the BiBoS Research Center (University of Bielefeld), 
the TMR network ERBFMRX-CT-970122 and the DFG under grant FOR 339/1-2 
are gratefully acknowledged.

\section{Appendix}
\setcounter{equation}{0}
The Euler-Poincar\'e characteristic has another equivalent
formulation showing closer connections with the topological properties of configurations.
FK bond configurations $X$, together with a suitably defined orientation,
can be endowed with a structure of $2$-dimensional (closed) cell-complexes
(see \cite{A,Wa3,L2}). Sites, bonds and plaquettes are then called,
respectively, $0$--cells, $1$--cells and $2$--cells. A cell complex is then
defined as a finite collection of $q$--cells, $q=0,1,\dots$

A theorem in Algebraic topology states that (see e.g. \cite{Ag}):
the only topologically invariant functions on cell complexes are those
which are functions of the Euler-Poincar\'e characteristic and the
dimension of the complexes.

Formal linear combinations of oriented $q$--cells are called $q$-chains and
form an additive group. Of primary importance is the definition of a
boundary map $\partial_{q}$ acting on a $q$-cell (more generally on
$q$-chains) giving rise to a $(q-1)$--cell, $q\ge 1$ (the boundary map of a
$0$--cell being defined as a $0$--cell). For example the boundary map
acting on an oriented plaquette ($2$--cell) leads to the set of all
oriented bonds ($1$-cells) defining its boundary.

Let $c^{(q)}$ a $q$--cell. If $\partial_{q}\, c^{(q)} =0$ then $c^{(q)}$ is
called a {\it $q$-cycle} and if $c^{(q+1)}$ is a $(q+1)$--cell such that
$\partial_{q+1} \, c^{(q+1)} = c^{(q)}$ then $c^{(q+1)}$ is called a {\it
$q$--boundary}. The set $Z_{q}$ of $q$-cycles and $B_{q}$ of
$q$--boundaries form subgroups of the group of $q$--chains. The quotient
group $H_{q} = Z_{q} / B_{q}$ is called the $q$-th homology group and it is
the topic of algebraic topology to show that the rank of these groups
(called the {\it Betti numbers}) is invariant under homotopy
transformations.

This means, loosely speaking, that there is a way to identify geometric
structures through continuous deformations without introducing or
suppressing ``holes' in them; for example a sphere in $\Reel^3$ is not
homotopy equivalent to a point. In our context of clusters of bond
configurations, a connected component (collection of plaquettes and bonds)
can be shrinked (and so it is homotopy equivalent) to a point. In that case
it turns out that the rank of the $0$-th homology group (the $0$-th Betti number)
has a very simple meaning, it just gives the number of connected components of the
configuration.

If $\b^{i}$ denotes the $i$--th Betti number, then for two dimensional cell complexes like the one we consider in this paper, one has
\begin{equation}
\chi(X) = \b^{0}(X) - \b^{1}(X)
\label{Betti}
\end{equation}
Here $\b^{1}(X)$ (the rank of the first homology group) is the number of holes in the configuration, i.e. cycles which do not surround a connected collection of plaquettes, each of them having 4 bonds on their boundary (see Fig. 8). One can verify that Eq. (\ref{Betti}) applies to all examples given in Fig. 1. 

In dimension $d=3$, $\b^{2}(X)$ would give the number of cavities inside the cell complex. 

\begin{figure}[htb]
\begin{center}
\epsfig{file=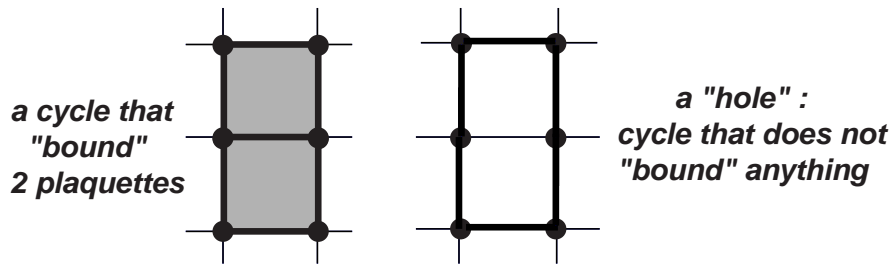,width=12cm}
\caption{Cycle and hole in FK-configuration.}
\end{center}
\end{figure}

Equation (\ref{Betti}) shows that the Euler-Poincar\'e characteristic of a bond configuration  depends only on homotopy equivalence classes of objects and as such, it is an
intrinsic property which provides a distinguished signature of its topological
features. The fact that it allows to probe the critical behaviour of physical systems makes possible the interpretation of phase transition on purely topological grounds. 

This formalism of cell complexes has been shown to be very useful in order
to define general duality arguments in several models of statistical
mechanics and it is the basis of the algebraic approach of the study of phase
transitions \cite{Wa3}.


\newpage
\begin{thebibliography}{99}

\bibitem[A]{A} P. S. Alexandroff, {\em Combinatorial Topology}, Graylock
Press, Rochester, (1956).

\bibitem[Ag]{Ag} M. K. Agoston, {\em Algebraic Topology}, Pure and Applied
Mathematics, M. Dekker
Inc. , New York, (1976).

\bibitem[AKPM]{AKPM} C. H. Arns, M. A. Knackstedt, W. V. Pinczewski, and K.
Mecke, Phys. Rev. E  {\bf 63}, 31112, (2001).

\bibitem[Ba]{Ba} I. Balberg, Phys. Rev. B {\bf 31}, 4053, (1985).

\bibitem[Bax]{Bax} R. J. Baxter, J. Phys. C {\bf 6}, L445, (1973), J. Phys.
A {\bf 15}, 3329, (1982)

\bibitem[Be]{Be} C. Berge, {\em Graphs and Hypergraphs}, North Holland,
(1976).

\bibitem[FK]{FK} C. M. Fortuin and P. W. Kasteleyn, Physica {\bf 57}, 536,
(1972).

\bibitem[Gra]{Gra} S. B. Gray, I.E.E.E. Trans. Comp. {\bf 20}, 551, (1971).

\bibitem[Gri]{Gri} G. Grimmett, {\em Percolation}, Springer, (1999).

\bibitem[Ha]{Ha} H. Hadwiger, {\em Vorlesungen \"uber Inhalt, Oberfl\"ache
und Isoperimetrie}, Springer, 1957.

\bibitem[HK]{HK} T. Hofs\"ass and H. Kleinert, J. Chem. Phys. {\bf 86},
3565, (1987).

\bibitem[Ho]{Ho} J. Hoshen, R. Kopelman, Phys. Rev. B {\bf 14}, 3438 (1976).

\bibitem[J]{J} J. P. Jernot and P. Jouannot, in: {\em Mathematical
Morphology and Applications to Image Processing. ISMM'94}, 35, (1994).

\bibitem[K]{K} R. Kotecky and S. Shlosman, Commun. Math. Phys. {\bf 83},
493, (1982).

\bibitem[KLMR]{L2} R. Kotecky, L. Laanait, A. Messager, J. Ruiz, J. Stat.
Phys. {\bf 58}, 199, (1990).

\bibitem[LMR]{L1} L. Laanait, A. Messager, and J. Ruiz, Commun. Math. Phys.
{\bf 105}, 527, (1986).

\bibitem[LMMRS]{L3} L. Laanait, S. Miracle-Sol\'e , A. Messager, J. Ruiz,
and S. Shlosman, Commun. Math. Phys. {\bf 140}, 81, (1991).


\bibitem[M1]{M1} K. R. Mecke, Applications of Minkowski Functionals in
Statistical Physics, in {\em Statistical Physics and Spatial Statistics.
The Art of Analyzing and Modeling Spatial Structures and Pattern
Formation}, Lecture Notes in Physics {\bf 554}, K. R. Mecke and D. Stoyan
Eds. Springer (2000).

\bibitem[M2]{M2} K. R. Mecke, {\em Complete Family of Fractal Dimensions
Based on Integral Geometry}, submitted to Physica A (2000).

\bibitem[MW]{MW} K. R. Mecke and H. Wagner, {\em Euler Characteristic and
Related Measures for Random Geometric Sets}, J. Stat. Phys. {\bf 64}, 843,
(1991).

\bibitem[O]{O} B. L. Okun, J. Stat. Phys. {\bf 59}, 523, (1990).

\bibitem[PS]{PS} G. E. Pike and C. H. Seager, Phys. Rev. B {\bf 10}, 1421,
(1974).

\bibitem[S]{S} L. A. Santal\`{o}, {\em Integral Geometry and Geometric
Probability}, Addison-Wesley, (1976).

\bibitem[Sa]{Sa} S. Fortunato and H. Satz, {\em Polyakov Loop Percolation
and Deconfinement in SU(2) Gauge Theory}, Phys. Lett. B {\bf 475}, 311 (2000);
S. Fortunato, F. Karsch, P. Petreczky, H. Satz, {\em Effective Z(2)
Spin Models of Deconfinement and Percolation in SU(2) Gauge Theory},
Phys. Lett. B {\bf 502}, 321 (2000).

\bibitem[Sch]{Sch} R. Schneider, {\em Convex Bodies: The Brunn-Minkowski
Theory}, Cambridge University Press, Cambridge (1993).

\bibitem[Se]{Se} J. Serra, {\em Mathematical Morphology}, Academic Press,
(1982).

\bibitem[SyEs]{SyEs} M. F. Sykes and J. W. Essam, Exact Critical
Percolation Probabilities for Site and Bond Problems in Two Dimensions, J.
Math. Phys. {\bf 5}, 1117-1121, (1964).

\bibitem[Wa1]{Wa1} H. Wagner, {\em Euler Characteristic for Archimedean
Lattices}, (2000). (unpublished)

\bibitem[Wa2]{Wa2} K. R. Mecke Th. Buchert and H. Wagner, Asrton.
Astrophys. {\bf 288}, 697, (1994).

\bibitem[Wa3]{Wa3} K. Dr\"{u}hl and H. Wagner, Ann. Phys. {\bf 141}, 225
(1982).

\bibitem[We]{We} F. J. Wegner, J. Math. Phys. {\bf 12}, 2259 (1971).
(1982).

\bibitem[Wo]{Wo} U. Wolff, Phys. Rev. Lett. {\bf 62}, 361 (1989).

\bibitem[Wu]{Wu} F. Y. Wu, Rev. Mod. Phys. {\bf 54}, 235 (1982).

\end{thebibliography}
\end{document}